\documentclass[conference]{IEEEtran}
\IEEEoverridecommandlockouts
\usepackage{cite}
\usepackage{amsmath,amssymb,amsfonts}
\usepackage{algorithmic}
\usepackage{graphicx}
\usepackage{textcomp}
\usepackage{xcolor}
\def\BibTeX{{\rm B\kern-.05em{\sc i\kern-.025em b}\kern-.08em
    T\kern-.1667em\lower.7ex\hbox{E}\kern-.125emX}}

\usepackage{import}
\usepackage{multicol}
\usepackage{multirow}
\usepackage{tikz}
\usepackage{amsmath}

\begin{document}

\def\edgecolor{rgb:blue,4;red,1;green,4;black,3}
\newcommand{\midarrow}{\tikz \draw[-Stealth,line width =0.8mm,draw=\edgecolor] (-0.3,0) -- ++(0.3,0);}

\title{Gradient Reduction Convolutional Neural Network Policy for Financial Deep Reinforcement Learning\\
\footnotesize  Regular Research Paper (CSCE-ICAI'24)}

\author{\IEEEauthorblockN{1\textsuperscript{st} Sina Montazeri}
\IEEEauthorblockA{\textit{Computer Science and Engineering} \\
\textit{University of North Texas}\\
Denton, United States \\
sinamontazeri@my.unt.edu}
\and
\IEEEauthorblockN{2\textsuperscript{nd} Haseebullah Jumakhan}
\IEEEauthorblockA{\textit{Artificial Intelligence Research Center (AIRC)} \\
\textit{Ajman University}\\
Ajman, United Arab Emirates \\
hjk608@nyu.edu}
\and
\IEEEauthorblockN{3\textsuperscript{nd} Sonia Afrasiabian}
\IEEEauthorblockA{\textit{Computer Science and Engineering} \\
\textit{University of North Texas}\\
Denton, United States \\
soniaafrasiabian@my.unt.edu}
\and
\IEEEauthorblockN{4\textsuperscript{rd} Amir Mirzaeinia}
\IEEEauthorblockA{\textit{Computer Science and Engineering} \\
\textit{University of North Texas}\\
Denton, United States \\
amir.mirzaeinia@unt.edu}
}
\maketitle

\begin{abstract}
Building on our prior explorations of convolutional neural networks (CNNs) for financial data processing, this paper introduces two significant enhancements to refine our CNN model's predictive performance and robustness for financial tabular data. Firstly, we integrate a normalization layer at the input stage to ensure consistent feature scaling, addressing the issue of disparate feature magnitudes that can skew the learning process. This modification is hypothesized to aid in stabilizing the training dynamics and improving the model's generalization across diverse financial datasets. Secondly, we employ a Gradient Reduction Architecture, where earlier layers are wider and subsequent layers are progressively narrower. This enhancement is designed to enable the model to capture more complex and subtle patterns within the data, a crucial factor in accurately predicting financial outcomes. These advancements directly respond to the limitations identified in previous studies, where simpler models struggled with the complexity and variability inherent in financial applications. Initial tests confirm that these changes improve accuracy and model stability, suggesting that deeper and more nuanced network architectures can significantly benefit financial predictive tasks. This paper details the implementation of these enhancements and evaluates their impact on the model's performance in a controlled experimental setting.
\end{abstract}


\section{Introduction}
As financial markets evolve with increasing complexity and data availability, applying machine learning techniques has shown promising results in navigating these dynamic environments. Building upon our prior works, which demonstrated the utility of CNNs in financial deep reinforcement learning (DRL) using an environment with continuous action spaces and later on strategically arranging feature vectors, we now aim to introduce further refinements. These refinements are designed not only to enhance model robustness and predictive performance but also to provide practical solutions for the challenges posed by financial datasets. These refinements are aimed at not just enhancing model robustness and predictive performance, but also at providing practical solutions for the challenges posed by financial datasets.

Financial datasets pose unique challenges due to their inherent volatility, non-stationarity, and the diversity of scale among different financial metrics, which necessitates innovative approaches in model architecture and preprocessing techniques. These challenges often lead to difficulties in model convergence and generalization, necessitating innovations in model architecture and preprocessing techniques. To address these challenges, our current research incorporates two significant enhancements into the CNN architecture specifically designed for financial tabular data: normalization of inputs and increasing the depth and breadth of the convolutional layers.

These enhancements aim to stabilize the training process and improve the model's ability to effectively extract and utilize complex patterns within the data. The subsequent sections will explore the mathematical foundations supporting these enhancements, underscoring their expected impact on the model's performance across varied financial datasets.

\section{Problem Description}
In this section, we outline our goal of improving our previous stock trading model, initially designed to maximize anticipated returns\cite{Montazeri2024a}.

Our new trading model also utilizes the FinRL-Meta environment, a platform designed to simulate dynamic market conditions for financial reinforcement learning (FinRL) by Liu et al. \cite{FinRL_Meta}. This environment addresses key challenges such as the low signal-to-noise ratio in financial data, survivorship bias, and model overfitting. FinRL-Meta transforms real-world financial data into standardized gym-style environments, enabling the training and evaluation of deep reinforcement learning (DRL) agents in conditions that closely mirror actual market behavior.

\subsection{Market Environment}
To model the financial market environment, FinRL-Meta structures its market environment as a Markov Decision Process (MDP), defined by the tuple \( (S, A, R, P, \gamma) \). This framework includes a comprehensive state space \( S \) that encapsulates the market's current status, an action space \( A \) that lists all possible trading actions, a reward function \( R \) that quantifies the results of these actions, transition probabilities \( P \) that model market dynamics, and a discount factor \( \gamma \) for future rewards. This MDP formulation provides a robust and flexible model for sequential decision-making in financial markets.

An MDP serves as a fundamental framework for modeling decision-making problems where outcomes are partly random and partly under the control of a decision-maker. It is formally defined by the tuple \( (S, A, R, P, \gamma) \), where \( S \) is the state space, \( A \) is the action space, \( R \) is the reward function, \( P \) is the state transition probability, and \( \gamma \) is the discount factor. In an MDP, the state transition probability \( P(s'|s, a) \) describes the likelihood of transitioning to state \( s' \) from state \( s \) when action \( a \) is taken. The reward function \( R(s, a, s') \) assigns a numerical value to the transition, quantifying the immediate benefit of taking action \( a \) in state \( s \) and arriving at state \( s' \). The discount factor \( \gamma \in (0, 1] \) determines the present value of future rewards, balancing immediate and long-term gains. The objective in an MDP is to find an optimal policy \( \pi^* \) that maximizes the expected cumulative discounted reward, expressed as:
\[ \pi^* = \arg\max_\pi \mathbb{E} \left[ \sum_{t=0}^{T} \gamma^t R(s_t, a_t, s_{t+1}) \right] \]
where \( T \) is the time horizon. This formulation is also used in FinRL which allows for a systematic approach to optimizing decision-making processes in complex, stochastic environments, such as financial markets.

\subsection{State Space}
Building on the MDP framework, the state space 
\( S \) in the FinRL-Meta environment captures essential market conditions and asset attributes required for effective trading decisions. Each state \( s_t \) at time \( t \) includes a combination of balance \( b \), asset prices \( p \in \mathbb{R}_+^{29} \), holdings \( h \in \mathbb{Z}_+^{29} \), and various technical indicators.

As detailed in Table \ref{tab:featurevectore} the daily feature vector used in the state representation includes the following features for a set of companies:

\begin{itemize}
    \item \textbf{Open}: The opening price of the asset for the trading day.
    \item \textbf{High}: The highest price of the asset during the trading day.
    \item \textbf{Low}: The lowest price of the asset during the trading day.
    \item \textbf{Close}: The closing price of the asset for the trading day.
    \item \textbf{Volume}: The total number of shares traded during the trading day.
    \item \textbf{Day}: The day of the week, providing a temporal context.
    \item \textbf{MACD (Moving Average Convergence Divergence)}: A trend-following momentum indicator that shows the relationship between two moving averages of an asset's price.
    \item \textbf{Bollinger Bands Upper Bound (Boll\_UB)}: The upper band of the Bollinger Bands, indicating overbought conditions.
    \item \textbf{Bollinger Bands Lower Bound (Boll\_LB)}: The lower band of the Bollinger Bands, indicating oversold conditions.
    \item \textbf{RSI (Relative Strength Index) 30}: A momentum oscillator that measures the speed and change of price movements, specifically over a 30-day period.
    \item \textbf{CCI (Commodity Channel Index) 30}: A momentum-based oscillator that measures the deviation of the asset's price from its statistical mean over a 30-day period.
    \item \textbf{DX (Directional Movement Index) 30}: An indicator used to assess the strength of a trend over a 30-day period.
    \item \textbf{Close 30 SMA (Simple Moving Average)}: The average of the closing prices over the last 30 days.
    \item \textbf{Close 60 SMA}: The average of the closing prices over the last 60 days.
    \item \textbf{VIX (Volatility Index)}: A real-time market index that represents the market's expectations for volatility over the coming 30 days.
    \item \textbf{Turbulence}: A measure of market stress and volatility, capturing unexpected and severe market movements.
\end{itemize}
\begin{table}[!htbp]
\begin{center}
\caption{Daily Feature Vector for 29 Companies}
\label{tab:featurevectore}
\begin{tabular}{|l|l|}
\hline
Name                         & Size \\ \hline
Balance                      & 1    \\ \hline
Open, High, Low, Close Prices & 29 each \\ \hline
Volume                       & 29   \\ \hline
Day                          & 29   \\ \hline
MACD                         & 29   \\ \hline
Bollinger Bands (Upper, Lower) & 29 each \\ \hline
RSI (30)                     & 29   \\ \hline
CCI (30)                     & 29   \\ \hline
DX (30)                      & 29   \\ \hline
SMA (30-day, 60-day)         & 29 each   \\ \hline
VIX                          & 29   \\ \hline
Turbulence                   & 29   \\ \hline
\end{tabular}
\end{center}
\end{table}

\begin{table}[!htbp]
\caption{Companies in Feature Vector}
\begin{center}
\begin{tabular}{|l|l|l|}
\hline
Company & Sector & Ticker \\ \hline
Apple Inc. & Technology & AAPL \\ \hline
Cisco Systems Inc. & Technology & CSCO \\ \hline
IBM Corp. & Technology & IBM \\ \hline
Intel Corp. & Technology & INTC \\ \hline
Microsoft Corp. & Technology & MSFT \\ \hline
Salesforce.com Inc. & Technology & CRM \\ \hline
Visa Inc. & Financials & V \\ \hline
Goldman Sachs Group Inc. & Financials & GS \\ \hline
JPMorgan Chase & Financials & JPM \\ \hline
American Express Co. & Financials & AXP \\ \hline
Travelers Companies Inc. & Financials & TRV \\ \hline
Amgen Inc. & Health Care & AMGN \\ \hline
Johnson \& Johnson & Health Care & JNJ \\ \hline
Merck \& Co. Inc. & Health Care & MRK \\ \hline
UnitedHealth Group Inc. & Health Care & UNH \\ \hline
Walmart Inc. & Consumer Staples & WMT \\ \hline
Procter \& Gamble Co. & Consumer Staples & PG \\ \hline
Coca-Cola Co. & Consumer Staples & KO \\ \hline
Walgreens Boots Alliance Inc. & Consumer Staples & WBA \\ \hline
Home Depot Inc. & Consumer Discretionary & HD \\ \hline
McDonald's Corp. & Consumer Discretionary & MCD \\ \hline
Nike Inc. & Consumer Discretionary & NKE \\ \hline
Walt Disney Co. & Consumer Discretionary & DIS \\ \hline
3M Co. & Industrials & MMM \\ \hline
Boeing Co. & Industrials & BA \\ \hline
Caterpillar Inc. & Industrials & CAT \\ \hline
Honeywell International Inc. & Industrials & HON \\ \hline
Chevron Corp. & Energy & CVX \\ \hline
Verizon Communications Inc. & Telecommunications & VZ \\ \hline
Dow Inc. & Materials & DOW \\ \hline
\end{tabular}
\label{tab:companylist}
\end{center}
\end{table}

\begin{figure}[!htbp]
    \vspace{-3pt} 
    \centering
    \begin{minipage}{0.9\columnwidth}
    \begin{tikzpicture}        
    
    \draw[step=0.5cm,gray,very thin] (0,0) grid (4,8);  
    \draw[thick] (0,0) rectangle (4, 8);               
    
    \node[anchor=east] at (0, 5) {1};                  
    \node[anchor=east] at (0, 3) {90};                 
    \node[anchor=east] at (0, 8) {2015/05/05};         
    \node[anchor=east] at (0, 0) {2023/05/01};         
    
    \node at (0.25, 7.75) {1};                         
    \node at (0.75, 7.75) {29};
    \node at (1.25, 7.75) {29};                        
    \node at (1.75, 7.75) {29};  
    \node at (2.25, 7.75) {...};                       
    \node at (2.75, 7.75) {...};  
    \node at (3.25, 7.75) {...};                       
    \node at (3.75, 7.75) {29}; 
    
    \draw[red, thick] (0,3) rectangle (4,5);           
    
    \small \node at (2, 8.25) {1 + (10 x 29 companies) = 291};               
    
    \end{tikzpicture}   
    \end{minipage}
    \caption{The Sliding Window to Create the Input Matrix for Our CNN}
    \label{fig:Sliding_window}
    \vspace{-3pt} 
\end{figure}
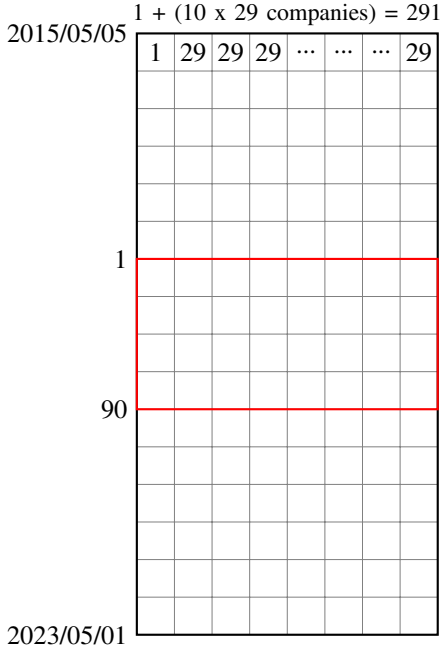

The resulting state representation is a high-dimensional vector reshaped into a 2D matrix format over a 90-day sliding window, as illustrated in Fig. \ref{fig:Sliding_window}. This method allows the CNN to effectively capture temporal patterns and relationships within the data.

\subsection{Action Space}
Following the state space, the action space 
\( A \) encompasses the possible trading actions the agent can perform, such as buying, selling, or holding stocks, directly affecting the portfolio's composition. Each action \( a_t \) at time \( t \) involves a vector representing the number of shares to buy or sell for each of the \( D \) stocks in the portfolio. The action space is continuous, allowing for fractional adjustments in holdings.

Actions are scaled to represent transaction magnitudes, ranging from -1000 to +1000 shares, with negative values indicating selling and positive values indicating buying. This scaling allows the agent to execute various trading strategies, from conservative adjustments to more aggressive trades, based on market conditions and the agent's learned policy. The flexibility in the action space enables the agent to adapt to different market scenarios, optimizing portfolio performance through informed trading decisions based on the current state.

\subsection{Reward Function}
As the final key element of our model framework, the reward function \( R(s, a, s') \) is designed to provide clear feedback on the agent's trading performance. It measures the change in portfolio value resulting from an action \( a \) taken in state \( s \), transitioning to a new state \( s' \). Formally, the reward is defined as:
\[ R(s_t, a_t, s_{t+1}) = v_{t+1} - v_t \]
where \( v_t \) and \( v_{t+1} \) represent the portfolio values at times \( t \) and \( t+1 \), respectively. This reward structure incentivizes the agent to make profitable trading decisions and manage risk effectively, aligning its actions with the overarching goal of maximizing long-term returns.

\section{Hypothesis}
Given our goal of enhancing the performance of Convolutional Neural Networks (CNNs) in financial deep reinforcement learning (DRL), we focus on two key areas: input normalization and network architecture expansion. Our hypothesis is rooted in the challenges posed by financial data, such as varying scales of financial indicators and the need for sophisticated pattern recognition. By integrating a normalization layer and increasing the depth and width of the CNN, we aim to stabilize the training process and improve the model's ability to capture complex financial patterns.

\subsection{Column-wise Normalization of Input Signals}
The introduction of a normalization layer within our CNN architecture is based on the observation that financial data, where features such as price scales and trading volumes can vary widely, potentially lead to gradient instability and slow convergence. Standardized inputs facilitate a more stable and efficient learning process. Therefore, we hypothesize that normalizing each column (feature) independently within the input data will facilitate more stable and effective model learning. Column-wise normalization addresses the issue where different financial indicators, such as trading volume and price changes, vary widely in magnitude and volatility. Each column is standardized to have zero mean and unit variance, calculated as:
\[
\text{Normalized}_{x_i} = \frac{x_i - \mu_i}{\sigma_i + \epsilon}
\]
where \(x_i\) is the \(i\)-th column of the input, \(\mu_i\) and \(\sigma_i\) are the mean and standard deviation of that column, and \(\epsilon\) is a small constant added to prevent division by zero. 

This method ensures that each feature contributes proportionately to the learning process, preventing any single feature from dominating due to its scale.

\subsection{Gradient Reduction Architecture}
The second enhancement over our previous CNN architecture involves structuring the network in a gradient reduction architecture, where the earlier layers are wider, capturing a broad range of features, and the subsequent layers are progressively narrower, distilling these features into more complex representations \cite{LeCun1998}. As described in our prior research, CNNs are inherently well-suited to recognizing and processing spatial hierarchies in data. This characteristic makes them particularly valuable for financial applications where input data can often display complex and hierarchical dependencies that are not immediately apparent. In financial DRL, each layer of a CNN can be understood as a feature detector that interprets various aspects of the input data, such as trends, anomalies, or cyclical patterns inherent in financial time series. 

To better demonstrate the reasoning behind our hypothesis, let us review the fundamental operations that occur during the learning process. Each convolutional layer in a CNN applies a set of learnable filters to the input feature map. The convolution operation for a single filter can be mathematically represented as:
\[
y_{ijk} = \sum_{m}\sum_{n}\sum_{o} W_{mno} \cdot x_{(i+m)(j+n)(k+o)} + b
\]
where:
\begin{itemize}
    \item \(y_{ijk}\) is the output of the convolution at position \((i, j, k)\),
    \item \(W_{mno}\) are the weights of the filter,
    \item \(x_{(i+m)(j+n)(k+o)}\) represents the input feature map,
    \item \(b\) is the bias,
    \item \(m, n, o\) iterate over the filter dimensions and channels.
\end{itemize}

This operation enables the detection of features regardless of their position in the input\cite{Fukushima1980} which is supporting our hypothesis that CNNs are especially suitable for financial data where important features might occur at different times.

Following each convolution operation, a non-linear activation function is applied. The Rectified Linear Unit (ReLU) is commonly used due to its effectiveness and simplicity:
\[
f(z) = \max(0, z)
\]
ReLU introduces non-linearity into the model, allowing the network to learn complex patterns. It is preferred over other activations like sigmoid or tanh primarily because it helps in alleviating the vanishing gradient problem, which is crucial for training deep networks effectively.

As the network depth increases, lower layers learn basic patterns which are then combined by higher layers to form more abstract features. This hierarchical learning can be described as a series of transformations:
\[
x^{(l+1)} = f(W^{(l)} * x^{(l)} + b^{(l)})
\]

where \(x^{l}\) and \(x^{l+1}\) are the input and output of the \(l\)-th layer, respectively. Each layer captures different levels of abstraction, which is essential for interpreting complex financial data where high-level features might depend on the subtle interactions of lower-level features.

When this operation is considered in the context of DRL, the state representation learned by the CNN significantly affects the policy and value function estimates. A deeper CNN can therefore develop a more nuanced understanding of the state space, which directly impacts the DRL agent’s ability to evaluate and choose between different actions based on potential future rewards. 

Therefore we hypothesize that the gradient reduction architecture ensures that the policy derived from these representations is capable of discerning and reacting to complex financial scenarios, potentially leading to more profitable decision-making strategies.

\section{Methodology}
The enhanced CNN architecture was implemented to evaluate its efficacy in a Deep Reinforcement Learning (DRL) environment, utilizing the Proximal Policy Optimization (PPO) algorithm. We will describe the detailed methodology, beginning by how PPO works and how our CNN network architecture is integrated into it.

\subsection{Proximal Policy Optimization (PPO)}

As mentioned before, we are utilizing the state-of-the-art reinforcement learning algorithm developed by OpenAI. As proposed by Schulman et al. \cite{PPO}, PPO is designed to address several limitations of earlier policy optimization methods, such as the instability and inefficiency of vanilla policy gradients and the complexity of Trust Region Policy Optimization (TRPO). There are several key concepts and mechanisms that make PPO effective in Financial Deep Reinforcement Learning:

\begin{enumerate}
    \item \textbf{Policy Gradient Methods}:
    At the core of PPO lies the concept of policy gradient methods, which directly parameterize the policy using a set of parameters $\theta$. The goal is to optimize these parameters to maximize the expected return from the policy. The optimization process involves computing the gradients of the policy performance with respect to $\theta$ and updating the parameters in a direction that increases expected returns. The policy gradient theorem provides the mathematical foundation for this process. Specifically, the gradient of the expected return $J(\theta)$ with respect to the policy parameters $\theta$ can be expressed as:
    \[
    \nabla_{\theta} J(\theta) = \mathbb{E}_{\tau \sim \pi_{\theta}} \left[ \sum_{t=0}^{T} \nabla_{\theta} \log \pi_{\theta}(a_t | s_t) \hat{A}_t \right]
    \]
    In this equation, $\tau$ represents a trajectory, which is a sequence of states, actions, and rewards. The policy $\pi_{\theta}(a_t | s_t)$ denotes the probability of taking action $a_t$ given state $s_t$ under the policy parameterized by $\theta$. The term $\hat{A}_t$ is the advantage function, which measures how much better an action is compared to a baseline (typically the value function).
    
    \item \textbf{Clipped Surrogate Objective}:
    To enhance the stability of the policy updates, PPO introduces a clipped surrogate objective function. This mechanism is designed to prevent excessively large updates to the policy, which could destabilize the training process. The key idea is to ensure that the new policy does not deviate too much from the old policy by clipping the probability ratio between the new and old policies. The probability ratio, $r_t(\theta)$, is defined as:
    \[
    r_t(\theta) = \frac{\pi_{\theta}(a_t | s_t)}{\pi_{\theta_{\text{old}}}(a_t | s_t)}
    \]
    The clipped objective function is then given by:
    \[
    L^{\text{CLIP}}(\theta) = \mathbb{E}_t \left[ \min \left( r_t(\theta) \hat{A}_t, \text{clip}(r_t(\theta), 1 - \epsilon, 1 + \epsilon) \hat{A}_t \right) \right]
    \]
    The clipping operation ensures that $r_t(\theta)$ remains within the range $[1-\epsilon, 1+\epsilon]$, where $\epsilon$ is a hyperparameter. This prevents large deviations that could destabilize the training process.
    
    \item \textbf{Advantage Function}:
    A critical component in PPO is the advantage function $\hat{A}_t$, which plays a vital role in reducing the variance of the policy gradient estimates, leading to more stable and efficient training. The advantage function measures the relative value of an action compared to a baseline, usually the value function. It is typically estimated using Generalized Advantage Estimation (GAE), which balances bias and variance through a parameter $\lambda$. The advantage function is calculated as:
    \[
    \hat{A}_t = \delta_t + (\gamma \lambda) \delta_{t+1} + \cdots + (\gamma \lambda)^{T-t+1} \delta_{T-1}
    \]
    where $\delta_t$ is the temporal difference error defined as:
    \[
    \delta_t = r_t + \gamma V(s_{t+1}) - V(s_t)
    \]
    Here, $r_t$ is the reward received at time step $t$, $\gamma$ is the discount factor, $V(s_t)$ is the value of state $s_t$, and $V(s_{t+1})$ is the value of the next state $s_{t+1}$.
    
    \item \textbf{Multiple Epochs of Updates}:
    Another significant feature of PPO is the ability to perform multiple epochs of minibatch updates using the same set of data. Unlike traditional policy gradient methods that perform a single update per data sample, PPO improves sample efficiency by iterating over the collected data multiple times. This approach makes better use of the data and ensures more thorough optimization, leading to improved policy performance.
    
    \item \textbf{First-order Optimization}:
    Finally, PPO employs first-order optimization methods, such as stochastic gradient descent (SGD) or the Adam optimizer, to update the policy parameters. First-order methods rely on the gradient of the objective function to update the model parameters. In the context of PPO, the objective function is the clipped surrogate objective, $ L^{\text{CLIP}}(\theta) $. The update rule for first-order optimization methods is given by:
    \[
    \theta_{k+1} = \theta_k + \alpha \nabla_{\theta} L^{\text{CLIP}}(\theta_k)
    \]
    In this rule, $\theta_k$ represents the parameters at iteration $k$, $\alpha$ is the learning rate, and $\nabla_{\theta} L^{\text{CLIP}}(\theta_k)$ is the gradient of the objective function with respect to the parameters.
    
    \begin{itemize}
        \item \textbf{Stochastic Gradient Descent (SGD)} updates the parameters by computing the gradient of the objective function on a mini-batch of data points. This introduces stochasticity into the updates, helping the optimization process to escape local minima and converge to a better solution. The update rule for SGD can be expressed as:
        \[
        \theta_{k+1} = \theta_k + \alpha \nabla_{\theta} L^{\text{CLIP}}(\theta_k)
        \]
        
        \item \textbf{Adam Optimizer} by Kingma and Ba \cite{adam}, is an adaptive learning rate optimization algorithm that combines the advantages of AdaGrad and RMSProp. Adam adjusts the learning rate for each parameter based on estimates of the first and second moments of the gradients. This helps in stabilizing and speeding up the convergence process. Adam maintains two moving averages for each parameter: the first moment estimate (mean) and the second moment estimate (uncentered variance). These estimates are used to scale the gradient updates adaptively:
        \[
        \theta_{k+1} = \theta_k + \alpha \frac{\hat{m}_k}{\sqrt{\hat{v}_k} + \epsilon}
        \]
        In this equation, $\hat{m}_k$ and $\hat{v}_k$ are the bias-corrected first and second moment estimates, respectively, and $\epsilon$ is a small constant added for numerical stability. The adaptive adjustment of the learning rate for each parameter helps in handling sparse gradients and noisy updates more effectively.
    \end{itemize}
\end{enumerate}

\subsection{Implementation in Stable-Baselines3}

In our experimentation, we utilized an implementation of PPO provided by Stable-Baselines3, a popular library for reinforcement learning in Python by Raffin et al. \cite{StableBaselines3} Stable-Baselines3 offers reliable and efficient implementations of various RL algorithms, including PPO, and is well-suited for research and application in diverse environments.

The PPO algorithm in Stable-Baselines3 is integrated with our enhanced CNN architecture to optimize the trading policy in the FinRL-Meta environment. The key steps in our PPO implementation are as follows:

\begin{enumerate}
    \item \textbf{Policy Initialization}:
    We initialize the policy network using our proposed CNN architecture, which processes the high-dimensional state representations from the FinRL-Meta environment. This initialization sets the stage for the learning process by defining the initial policy.
    
    \item \textbf{Data Collection}:
    The agent interacts with the environment to collect trajectories, which are sequences of states, actions, rewards, and next states. These trajectories form the basis for estimating the advantage function and computing the surrogate objective.
    
    \item \textbf{Optimization}:
    Using the collected trajectories, multiple epochs of minibatch updates are performed on the policy network. The clipped surrogate objective is optimized using the Adam optimizer, ensuring stable and efficient updates.
    
    \item \textbf{Policy Update}:
    The policy parameters are updated based on the gradients computed from the surrogate objective. The clipping mechanism ensures that the updates remain within a safe range, preventing large deviations from the current policy.
    
    \item \textbf{Iteration}:
    Steps 2-4 are repeated iteratively until the policy converges or the maximum number of training steps is reached.
\end{enumerate}

\subsection{Experimental Validation}
To validate the effectiveness of PPO with our enhanced CNN architecture, we conducted extensive experiments in the FinRL-Meta environment. The performance of the PPO agent was compared against baseline models, including a multi-layer perceptron (MLP) and our previous CNN-based policy. The results demonstrated that the combination of PPO and our enhanced CNN architecture significantly outperformed the baselines in terms of cumulative rewards and stability.

In conclusion, PPO provides a robust and efficient framework for optimizing policies in complex environments like financial markets. Its integration with our enhanced CNN architecture leverages the strengths of both approaches, resulting in a powerful tool for financial deep reinforcement learning.

\subsection{Network Architecture}
In our previous work, we developed a Convolutional Neural Network (CNN) designed to operate on matrix state representation (2D) datasets, such as images and videos, learning features through the optimization of filters (kernels). This network consisted of Convolutional, Pooling, Normalization, Dropout, and Fully Connected layers. However, training deep networks posed challenges due to issues like vanishing and exploding gradients, which we addressed by employing regulated weights and batch normalization across layers.

The convolutional layers, with varying kernel sizes and strides, were specifically designed to capture temporal patterns and correlations within financial data, crucial for stock market data where the interplay of factors like stock prices, trading volumes, and technical indicators is complex and dynamic. Batch Normalization following each convolutional layer helped stabilize the learning process by normalizing the input to each layer, essential given the variability and non-stationarity of financial data.

To employ 2D CNNs in financial data analysis, we restructured the representation of our environment state into a matrix format. Financial data inherently manifests as a vector composed of daily features, where attributes like price are updated daily, while others, such as revenue and assets, update quarterly. The matrix representation concatenated a 90-day feature vector to create a matrix state representation, facilitating the analysis of temporal patterns and correlations in financial data.

In our current model, we continue the approach of arranging input features to ensure related features are positioned adjacent to each other, enhancing the network's ability to extract meaningful patterns \cite{Montazeri2024b}. For example, placing the closing prices and number of shares for each ticker close to each other allows the network to learn relationships between these variables more effectively. This preprocessing step facilitates efficient learning and contributes to the robustness and generalizability of the model across diverse financial datasets.

Previously, our CNN architecture included convolutional layers designed to capture temporal patterns and correlations within financial data. The initial convolutional layer used 32 filters with a kernel size of 8 and stride of 4, followed by ReLU activation and dropout. Subsequent layers included a second convolutional layer with 64 filters, also followed by ReLU activation and dropout. This structure allowed the network to capture basic features but was limited in its ability to extract more complex interactions due to the simpler architecture.

\begin{figure}[htbp]
    \centering
    \includegraphics[width=1\linewidth]{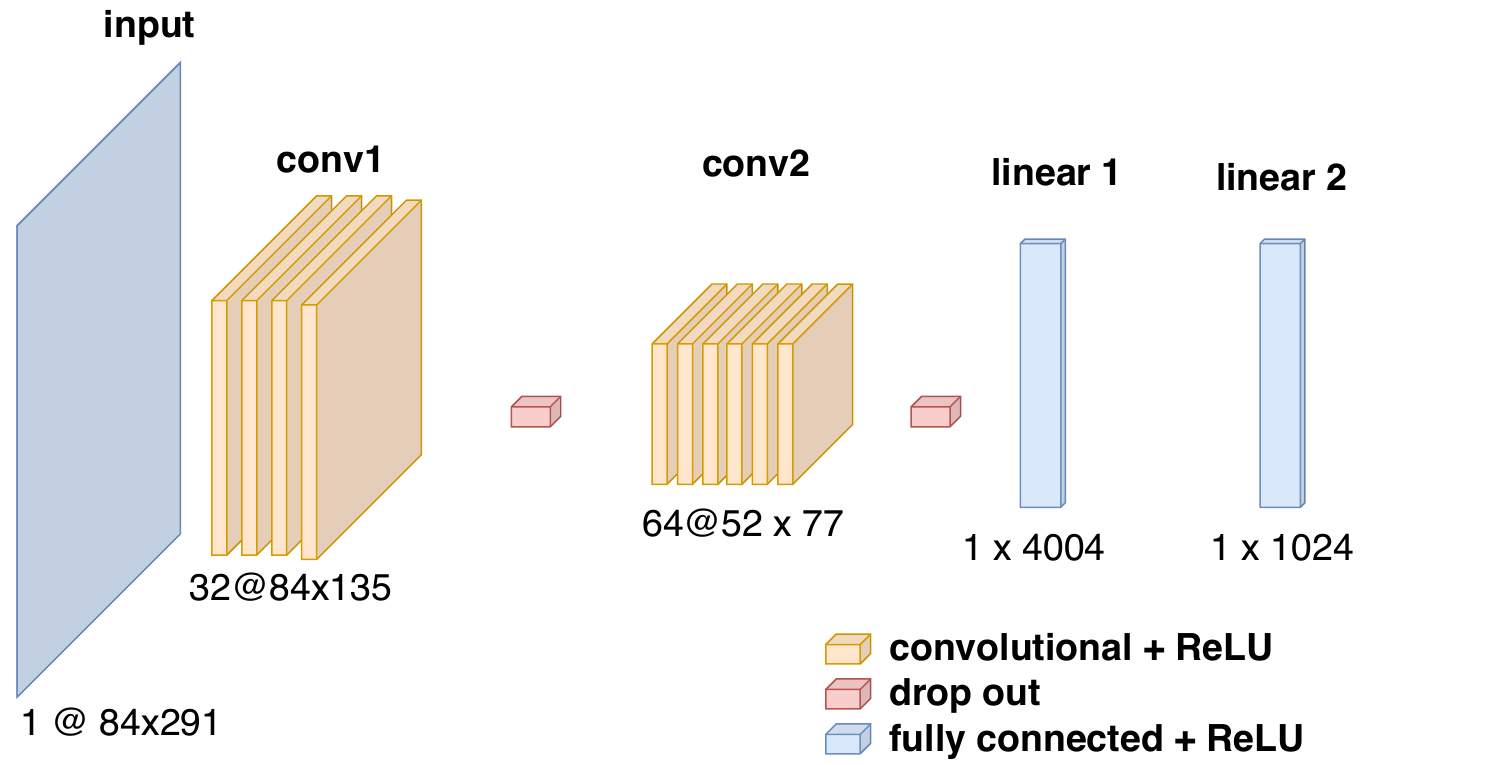}
    \caption{Original CNN Architecture}
    \label{fig:enter-label}
\end{figure}

In our new architecture, we have enhanced these layers to better optimize filter parameters and address the complexities of financial data. The initial convolutional layer now uses 32 filters with a kernel size of 8 and stride of 4, followed by Batch Normalization, ReLU activation, and Max Pooling. Subsequent layers have been made more complex, with the second convolutional layer using 64 filters, and later layers using 128 and 256 filters with adjusted kernel sizes and strides. These layers capture a broader spectrum of features, from basic price movements to complex interactions between different financial indicators. The increased size and depth of the model align with our hypothesis that a deeper network can better capture complex financial patterns, leading to improved policy derivation for financial decision-making. Additionally, the inclusion of batch normalization and max pooling in the new design stabilizes the learning process and improves feature extraction efficiency.

\begin{figure}[htbp]
    \centering
    \includegraphics[width=1\linewidth]{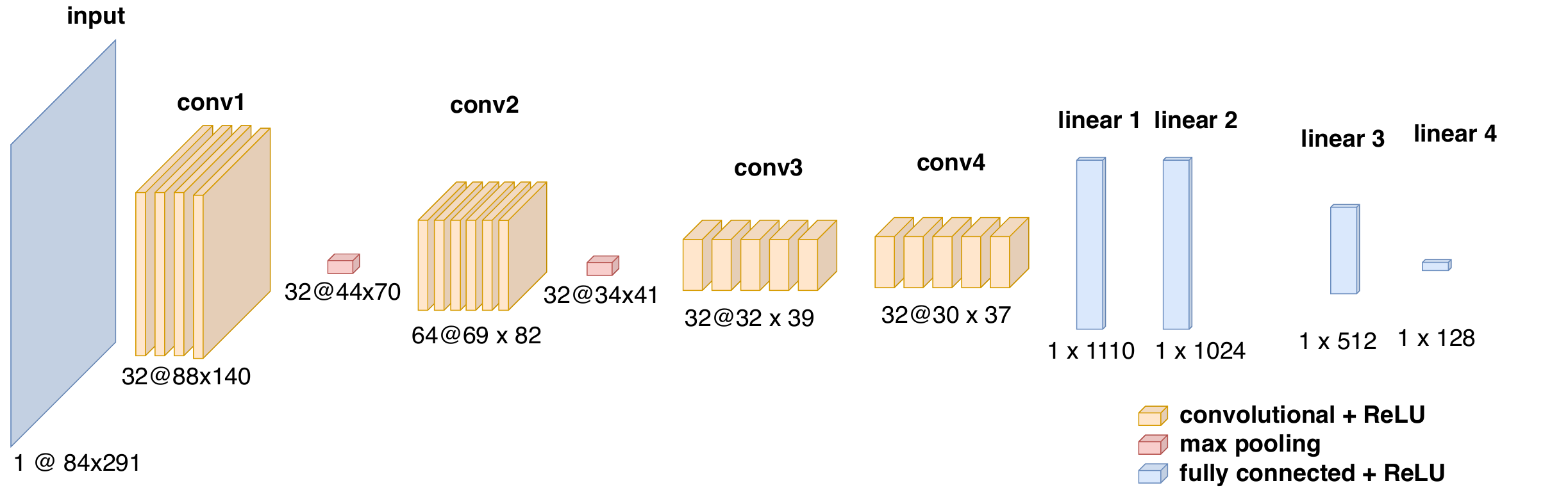}
    \caption{New Proposed Architecture}
    \label{fig:enter-label}
\end{figure}

\begin{figure*}[!htbp]
    \centering
    \includegraphics[width=1\linewidth]{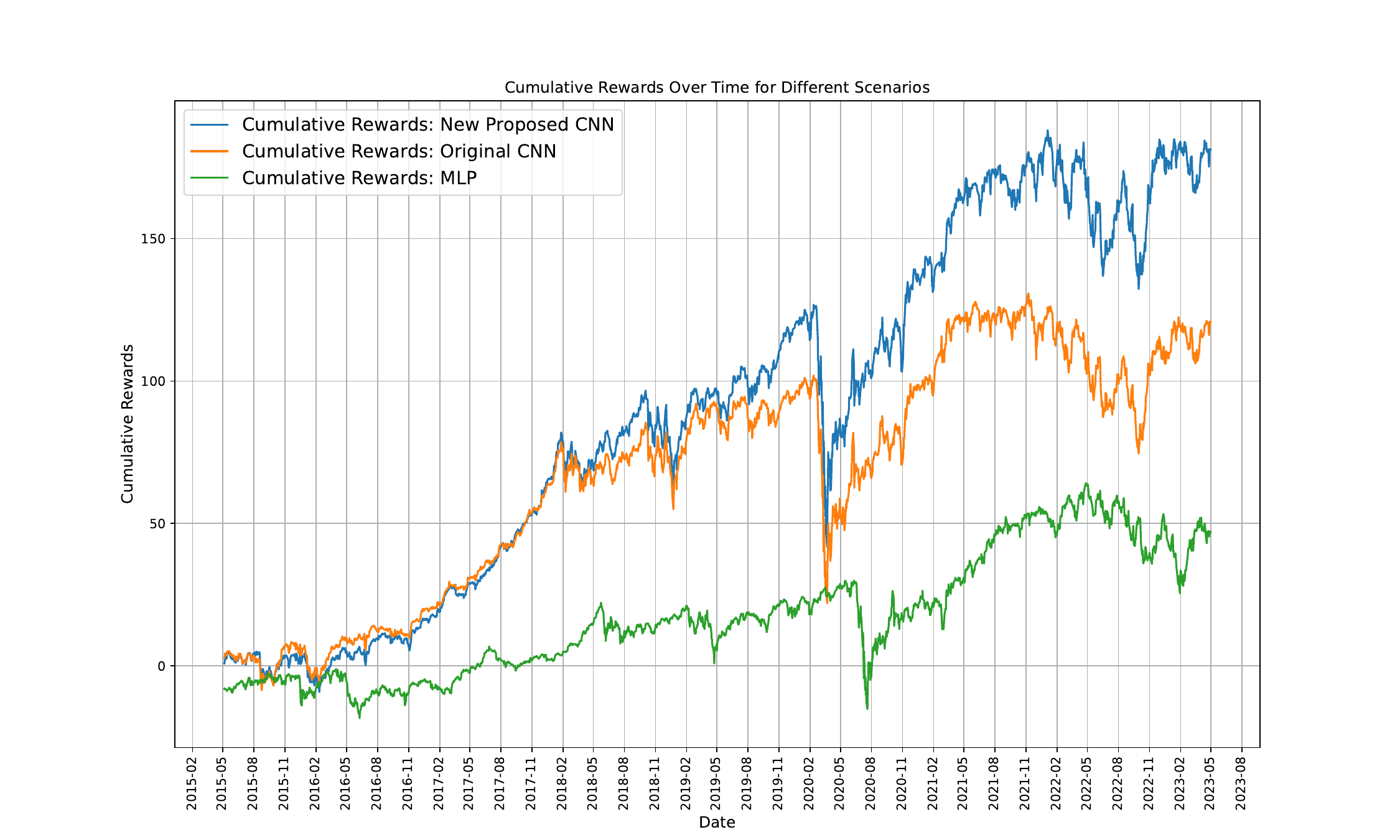}
    \caption{Commutative Rewards Comparison}
    \label{fig:results}
\end{figure*}

To validate our proposed changes, we conducted a comparative study. Models, including a multi-layer perceptron, our previous CNN, and the new proposed CNN, were trained under the same DRL framework to ensure consistent comparison, with all parameters kept constant and seeded. The rigorous training process was followed by a performance evaluation, where models were assessed based on their ability to maximize cumulative rewards in a simulated trading environment. This phase tests the practical utility of each model and demonstrates their effectiveness in real-world trading scenarios, as cumulative rewards reflect the soundness of trading decisions influenced by the model’s predictions.

\subsection{Model Training}
Our study involves training three distinct models: a Multi-layer Perceptron (MLP), our previous CNN, and the newly proposed CNN. To ensure consistency and fairness in comparison, we utilized the same environment with identical seeds for all probabilistic operations. The training parameters are kept constant and the mentioned seeds are added to maintain uniformity fairness of the challenge for each model.

During its run, each model processes financial data represented as a multidimensional array, where each feature corresponds to a different financial indicator or technical attribute of the market. The input data is preprocessed to fit the model specifications, including normalization and feature rearrangement. The training process involves feeding these preprocessed inputs into the models, which then learn to make trading decisions based on the patterns and correlations they detect within the data. The training phase is iterative, with the models continually adjusting their weights and biases to minimize the error in their predictions and improve their decision-making capabilities.

\subsection{Performance Evaluation}
After training, the models are evaluated based on their ability to maximize cumulative rewards in a simulated trading environment. This evaluation method provides a practical measure of each model's effectiveness in real-world trading scenarios. The cumulative rewards are calculated by summing the rewards obtained in each episode of the simulation, reflecting the success of the trading decisions made by the models.

Since in this simulated environment, each model operates as an agent making trading decisions based on the input financial data, the reward function is designed to reflect the profitability of these decisions. This incentivizes actions that lead to higher returns. By comparing the cumulative rewards, we can assess the relative performance of each model. This comparison helps in determining the improvements brought by the new CNN architecture over the previous models and provides insights into their potential applicability in live trading environments.

\section{Results}

The effectiveness of the gradient reduction architecture is demonstrated by comparing its performance with that of our original CNN model and a baseline MLP model. The scaled rewards were created by multiplying the difference of remaining assets and initial assets at the end of trading day by a scaling factor of 1e-4 and later used to compare the performance of the model.

\begin{itemize}
    \item \text{MLP Model}: Achieved a cumulative reward of 47.
    \item \text{Original CNN Model}: Showed improved performance with a cumulative reward of 120.
    \item \text{Gradient Reduction CNN}: Markedly outperformed the other models with a cumulative reward of 181.
\end{itemize}

These results, depicted in Figure \ref{fig:results}, indicate that the modifications to the CNN architecture significantly enhance its capability to interpret and predict complex financial data patterns more effectively than simpler models.

The performance comparison over time highlights how each model responded to various financial events between 2015 and 2023. During periods of market volatility, such as the 2016 Brexit vote and the 2020 COVID-19 pandemic, the new proposed CNN demonstrated superior adaptability and resilience. 

For instance, the original CNN and MLP models experienced significant fluctuations in cumulative rewards during these events, reflecting their struggle to adapt to sudden market changes. In contrast, the new CNN model showed a more stable and upward trajectory, indicating its enhanced ability to capture and respond to complex financial patterns. This stability can be attributed to the deeper and wider architecture, which allows the model to detect subtle interactions between financial indicators and make more informed trading decisions. Moreover, during the 2018 market corrections and the subsequent recovery periods, the new CNN maintained a higher cumulative reward compared to the other models. This suggests that the enhancements in the new architecture significantly contributed to better generalization and robustness in unpredictable market conditions.

Overall, the cumulative rewards over time for the new proposed CNN illustrate its improved performance in handling diverse and complex financial scenarios. This aligns with our hypothesis that a gradient reduction network with column-wise normalization can better capture intricate financial patterns, leading to more effective policy derivation and decision-making in financial deep reinforcement learning.

\section{Conclusion}
In this paper, we have introduced significant enhancements to our convolutional neural network (CNN) model tailored for financial deep reinforcement learning (DRL) applications. By integrating a normalization layer at the input stage, we addressed the issue of disparate feature magnitudes, thereby stabilizing the training dynamics and improving model generalization. Additionally, by employing a Gradient Reduction Architecture, where earlier layers are wider and subsequent layers are progressively narrower, we enabled the model to capture more complex and subtle patterns within the financial data.

Our empirical results demonstrate that these enhancements lead to superior predictive performance and robustness compared to previous simpler models. The Gradient Reduction Architecture proved particularly effective in extracting meaningful patterns from the high-dimensional financial data.

Furthermore, the integration of the Proximal Policy Optimization (PPO) algorithm within our framework provided a robust mechanism for optimizing trading policies. The experimental validation using the FinRL-Meta environment confirmed that our enhanced CNN model significantly outperformed baseline models, achieving higher cumulative rewards and greater stability in diverse market conditions.

In conclusion, the proposed enhancements to the CNN architecture, combined with advanced DRL techniques, offer a powerful approach for financial data processing and predictive modeling. These advancements hold promise for developing more effective and resilient financial trading strategies, contributing to the broader field of financial machine learning.

\vspace{12pt}
\end{document}